\documentclass[shortnote,onecolumn]{jpsj3}
\usepackage{txfonts}
\usepackage{color}
\newcommand{\SVd}{Sr$_2$VFeAsO$_{3-\delta}$}

\newcommand{\SVdb}{Sr$_2$VFeAsO$_{3-\delta}$ ($\delta = 0.25$)}

\title{Local Magnetic States of the Weakly Ferromagnetic Iron-Based Superconductor \SVd\ Studied by X-ray Magnetic Circular Dichroism}

\author{Masafumi Horio$^1$, Yukiharu Takeda$^2$, Hiromasa Namiki$^3$, Takao Katagiri$^3$, Yuki K. Wakabayashi$^4$, Shoya Sakamoto$^1$, Yosuke Nonaka$^1$, Goro Shibata$^1$, Keisuke Ikeda$^1$, Yuji Saitoh$^2$, Hiroshi Yamagami$^2$, Takao Sasagawa$^3$, and Atsushi Fujimori$^1$}
\inst{$^1$Department of Physics, University of Tokyo, Bunkyo-ku, Tokyo 113-0033, Japan \\
$^2$Materials Sciences Research Center, Japan Atomic Energy Agency (JAEA), Sayo, Hyogo 679-5148, Japan \\
$^3$Laboratory for Materials and Structures, Tokyo Institute of Technology, Yokohama, Kanagawa, 226-8503, Japan \\
$^4$Department of Electrical Engineering and Information Systems, University of Tokyo, Bunkyo-ku, Tokyo 113-0033, Japan}

\abst{We have performed x-ray magnetic circular dichroism (XMCD) measurements on the iron-based superconductor \SVd\ to study the origin of weak ferromagnetism (WFM) reported for this compound. While Fe 3$d$ electrons show a magnetic response similar to the other iron pnictides, signals from V 3$d$ electrons remain finite at zero magnetic field and may be responsible for the WFM.}

\begin{document}
\maketitle
Iron-based superconductors have attracted extensive interest since their discoveries. Despite the lack of thorough understanding on the microscopic mechanism of Cooper pairing, there is a consensus that the proximity to antiferromagnetism plays a key role as in the cuprates \cite{Chubukov2012}. Among all the iron-based superconductors, \SVd\ is a rare case where ferromagnetism, which is usually destructive to superconductivity, emerges at low temperatures and co-exists with superconductivity \cite{Sefat2011,Tojo2018}. A small spontaneous ferromagnetic moment of $\sim 10^{-4} \mu_\mathrm{B}$ for the $\delta \sim 0$ compound \cite{Cao2010} is enhanced to  $10^{-3}$--$10^{-1} \mu_\mathrm{B}$ by the introduction of oxygen vacancies \cite{Sefat2011,Tojo2018}. The \SVd\ lattice consists of an alternate stack of FeAs and perovskite-like Sr$_2$VO$_3$ layers \cite{Zhu2009}. While the FeAs layer serves as a stage for high-temperature superconductivity below 37 K for $\delta \sim 0$ \cite{Zhu2009}, Mott insulating state has been proposed for V 3$d$ electrons in the Sr$_2$VO$_3$ layer \cite{Qian2011}.

The central issue for \SVd\ is therefore the nature of the magnetic states. Cao \textit{et al}. \cite{Cao2010} concluded from the Fe$^{57}$ M\"{o}ssbauer spectra that no magnetic transitions occur at Fe in the range of $T = 20$--300 K, and hence proposed that V gives rise to the weak ferromagnetism (WFM) through canting of the antiferromagnetically ordered moments \cite{Nakamura2010}. In contrast, an antiferromagnetic transition at 165 K was suggested at Fe sites from a nuclear magnetic resonance (NMR) study by Ueshima \textit{et al}. \cite{Ueshima2014} Ok \textit{et al.} \cite{Ok2017} instead ascribed the transition at 155 K to non-magnetic charge/orbital ordering of Fe and regarded the ferromagnetic signal as a side effect. Thus not only the origin of the WFM but also the nature of the magnetic state below the transition temperature remains unclear. In this short note, we apply x-ray magnetic circular dichroism (XMCD), which provides element-specific information about the spin and orbital magnetic moments. $\delta = 0.25$ samples with larger ferromagnetic moments than $\delta = 0$ were measured. The results suggest that the magnetic moments of V 3$d$ rather than Fe 3$d$ contribute to the WFM in \SVd.

Bulk single crystals of \SVdb\ were grown by the self-flux method. Superconducting and ferromagnetic transition temperatures were determined from magnetic susceptibility measurements using a SQUID magnetometer to be 20 K and 210 K, respectively. At 50 K, a remanence magnetization of $2 \times 10^{-3}\mu_\mathrm{B}$/f.u. was detected. X-ray absorption spectroscopy (XAS) and XMCD measurements were conducted at beamline BL23SU of SPring-8 in the total electron-yield (TEY) mode at $T = 50$ K. Magnetic fields of 0.2--10 T were applied parallel to the $c$-axis of the crystal. The samples were cleaved in vacuum ($< 1 \times 10^{-9}$ Torr) to obtain clean surfaces.

\begin{figure}
\begin{center}
\includegraphics[width=120mm]{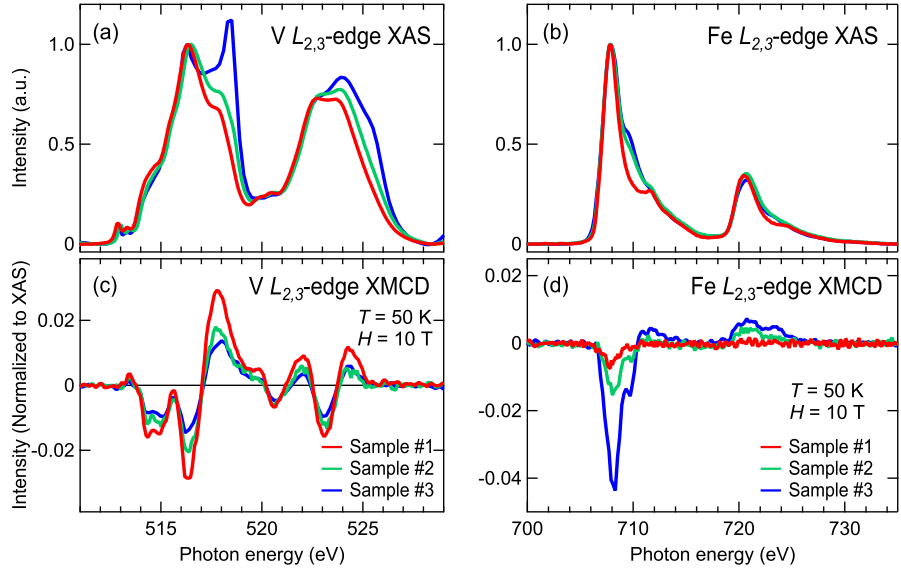}
\end{center}
\caption{(Color online) XAS and XMCD spectra for various \SVd\ samples. (a),(b) V and Fe $L_{2,3}$-edge XAS spectra, respectively, for samples \#1--\#3. The spectra have been normalized to the peak height at $\sim 516$ eV for the V edge and that at $\sim 708$ eV for the Fe edge. (c),(d) Corresponding XMCD spectra measured under $H = 10$ T at $T = 50$ K, and normalized to the XAS intensity.}
\label{XAS}
\end{figure}

Figures~\ref{XAS}(a) and (b) show V and Fe $L_{2,3}$-edge XAS spectra for different samples, i.e., different cleaved surfaces. The spectral line shape apparently depends on the sample. While the V XAS spectrum of sample \#1 resembles that of V$_2$O$_3$ \cite{Abbate1993}, consistent with dominant 3+ valence state reported in previous studies \cite{Cao2010}, additional spectral weight emerges at higher photon energies for samples \#2 and \#3, suggesting the existence of higher valence state such as V$^{5+}$. The spectrum gains weight at higher energies also for Fe XAS of the same samples. These results suggest that the cleaved surfaces on samples \#2 and \#3 are somewhat oxidized probably because the crystals originally had cracks and were cleaved along them. The oxidation also affects the magnetic states as evident from XMCD spectra [Figs.~\ref{XAS}(c) and (d)]. For V, the oxidation decreases the magnitude of the XMCD signal. This is consistent with the fact that V$^{5+}$ has no 3$d$ electrons and hence no magnetic moments. In contrast, the XMCD signal is enhanced for Fe in samples \#2 and \#3 probably due to the increase of ferromagnetic iron oxides. In order to discuss the intrinsic magnetic properties of \SVd, we will focus on the data for sample \#1 in the following.

\begin{figure}
\begin{center}
\includegraphics[width=120mm]{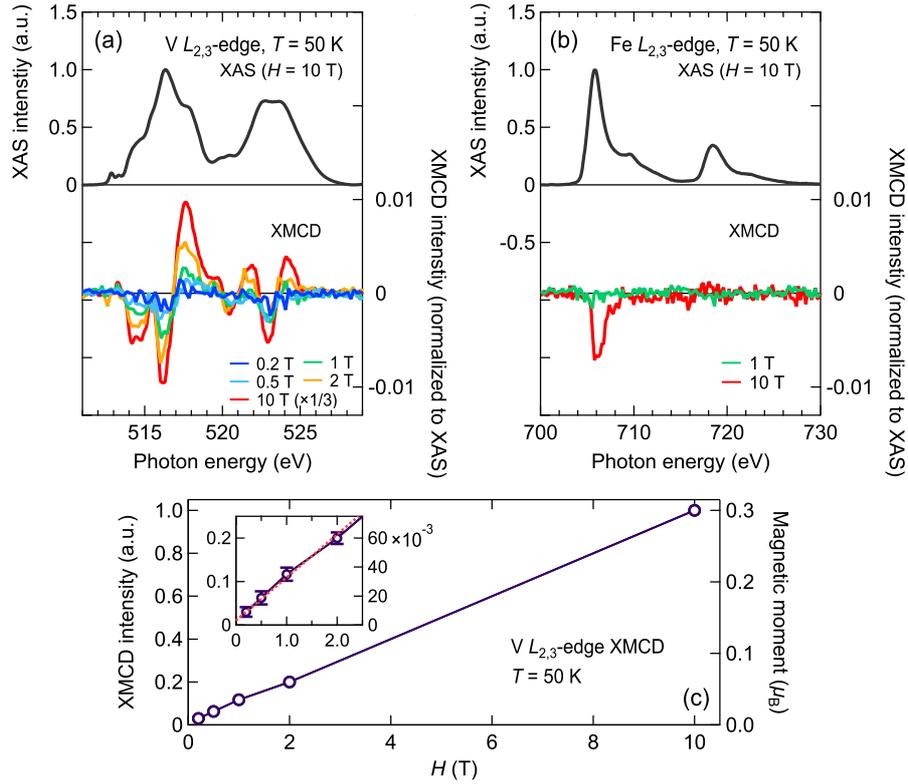}
\end{center}
\caption{(Color online) XMCD spectra acquired for sample \#1. (a),(b) V and Fe $L_{2,3}$-edge XAS (top) and XMCD (bottom) spectra, respectively. Note that the $H = 10$ T spectrum for V XMCD is multiplied by 1/3 for a display purpose. (c) V $L_{2,3}$-edge XMCD intensity versus magnetic field. The right axis indicates magnetic moment converted from the XMCD intensity. The inset shows a magnified plot at low magnetic fields and the extrapolation to $H = 0$ T (red dashed line).}
\label{XMCD}
\end{figure}

XMCD spectra obtained from sample \#1 with varying magnetic field are presented in Figs.~\ref{XMCD}(a) and (b). While finite XMCD is found at $H = 10$ T for Fe, the signal goes below the detection limit when the magnetic field is decreased down to 1 T. Applying the sum rule to the $H = 10$ T spectrum assuming the $d^6$ electron configuration, the spin, orbital, and total magnetic moments are deduced to be $1.2 \times 10^{-2}$, $5 \times 10^{-3}$, and $1.7 \times 10^{-2} \mu_\mathrm{B}$/ Fe atom, respectively, and the magnetic susceptibility of $1.9 \times 10^{-3}$ emu/Oe mol is obtained. The susceptibility value is of similar magnitude to BaFe$_2$As$_2$ or its Co-doped compounds ($\sim 5 \times 10^{-4}$ emu/Oe mol) \cite{Sefat2008}, and no signature of WFM is observed for Fe. This result is compatible with the previous M\"{o}ssbauer study \cite{Cao2010}, which reported no ordered moments for Fe, and also with the NMR study \cite{Ueshima2014}, where the antiferromagnetic transition similar to BaFe$_2$As$_2$ was observed.

As for V, the XMCD signal remains finite down to $H = 0.2$--0.5 T. Although the spin sum rule cannot be applied to light transition-metal elements such as V due to the overlap between the $L_2$ and $L_3$ edges, one can still discuss the magnetic moment from the magnitude of the XMCD signal. We first assumed that the XMCD line shape does not depend on magnetic field. This is reasonable since relevant parameters such as Coulomb interaction and transfer integrals should not change significantly by the application of magnetic field. Then, all the V XMCD spectra were fitted to the $H = 10$ T line shape with varying its magnitude. Thus extracted XMCD magnitude is plotted as a function of the magnetic field in Fig.~\ref{XMCD}(c) with an error bar defined as 2$\sigma$ of the fitting. The plot is almost linear, but extrapolation to $H = 0$ T (red dashed line) leaves finite XMCD signal. One method to obtain quantitative information about the V magnetic moment is comparison with configuration-interaction cluster calculation that yields the spin and orbital magnetic moments. In \SVd, oxygen atoms sit at octahedral sites around the V atoms with one apical oxygen atom missing. Still, the XAS and XMCD spectra are similar to what obtained for compounds where V atoms are octahedrally coordinated by O atoms \cite{Nonaka2018}. By comparing the present XMCD spectra with cluster-model calculation \cite{Nonaka2018}, we obtain an approximate value for the V magnetic moment of 0.3 $\mu_\mathrm{B}$ at $H = 10$ T. Then from Fig.~\ref{XMCD}(c), one finds the remnant magnetic moment of $3 (\pm 2) \times 10^{-3}$ $\mu_\mathrm{B}$, which is consistent with the present SQUID measurement. This implies that the V magnetic moments make a significant contribution to the WFM in \SVd.

Very recently, Tojo \textit{et al}.~\cite{Tojo2018} reported oxygen-vacancy dependent phase diagram of \SVd. They concluded from first-principles calculations that oxygen vacancies induce ferrimagnetic order of V and hence give rise to WFM, whereas their M\"{o}ssbauer measurements indicated that Fe is paramagnetic for superconducting compositions \cite{Tojo2018}. Our results are consistent with their study and further provide experimental and quantitative information about V magnetic moment. Canting of the antiferromagnetically ordered V moments \cite{Nakamura2010,Ueshima2014} also remains another possible explanation for the WFM.

In conclusion, we have conducted XMCD measurements on \SVdb\ and studied its magnetic states element-specifically. Magnetic susceptibility of Fe was found to be comparable to that of the other iron pnictides such as Ba(Fe,Co)$_2$As$_2$, and no signature of ferromagnetism was observed. One order of magnitude larger magnetic moment was detected for V, which appears to remain finite ($\sim 10^{-3} \mu_\mathrm{B}$) even under zero magnetic field. Our results are thus compatible with scenarios that V is the origin of the WFM.

\begin{acknowledgment}
This work was performed under the Shared Use Program of Japan Atomic Energy Agency (JAEA) Facilities (Proposal No. 2015A-E28) supported by JAEA Advanced Characterization Nanotechnology Platform as a program of gNanotechnology Platformh of the Ministry of Education, Culture, Science and Technology (MEXT) (Proposal No. A-15-AE-0026). The synchrotron radiation experiments were performed at the JAEA beamline BL23SU in SPring-8 (Proposal No. 2015A3881). The measurements using a SQUID magnetometer were made at the Cryogenic Research Center, the University of Tokyo. This work was supported by JSPS KAKENHI Grant Nos. 15H02109 and 14J09200. M.H., Y.K.W., and S.S. acknowledge support from the Program for Leading Graduate Schools. M.H. acknowledges support from the JSPS Fellowship for Young Scientists.
\end{acknowledgment}


\end{document}